\documentclass[aip,cha,reprint,showpacs,showkeys]{revtex4-1}
\usepackage{amssymb, dcolumn,  graphicx, latexsym, amsfonts, bm}

\begin{document}
\title{The ion acoustic instability of the rotating cylindrical helicon discharge plasma}
\author{V. V. Mikhailenko}\email[E-mail:]{vladimir@pusan.ac.kr}
\affiliation{BK21 FOUR Information Technology, Pusan National University,  Busan 46241, 
South Korea.}
\author{Hae June Lee}\email[E-mail:]{haejune@pusan.ac.kr}
\affiliation{Department of Electrical Engineering, Pusan National University, 
Busan 46241, South Korea}
\author{V. S. Mikhailenko}
\affiliation{Plasma Research Center, Pusan National University,  Busan 46241, South Korea}
\author{M. O. Azarenkov}
\affiliation{V.~N. Karazin Kharkiv National University, 
Kharkiv 61022, Ukraine}
\date{\today}

\begin{abstract} 
The kinetic theory for the cylindrical plasma, produced by the cylindrically 
symmetric (azimuthal mode number $m=0$) helicon wave, is developed with accounting for the 
cylindrical geometry and the radial inhomogeneity of the helicon wave and plasma. This theory
reveals macroscale effect of the azimuthal steady rotation of electrons with a radially 
inhomogeneous angular velocity, caused by radial inhomogeneity 
of the helicon electric field. It is found that this sheared rotation  as well as the electron 
density and temperature inhomogeneity are responsible for the development of the high 
frequency ion acoustic instability of the inhomogeneous cylindrical plasma. This instability is 
spatially localized in the region of strong gradients of the helicon wave electric field and 
of the plasma density, where it is more stronger than the parametric instabilities driven by the 
oscillating  motion of the electrons relative to ions in the helicon wave.
\end{abstract}

\maketitle

\section{Introduction}\label{sec1}
The helicon plasma sources attract great interest in plasma community and have various 
applications\cite{Shinohara, Takahashi} due to the remarkably strong absorption of helicon waves 
in plasmas and anomalously strong electron heating\cite{Boswell}. The helicon plasma 
is generated by using the helicon wave, which is the whistler wave in the bounded plasma. 
The frequency of helicon/whistler wave lies between ion and electron cyclotron frequencies. 
The linear theory of the helicon wave 
predicts that this wave has phase velocity much above the electron thermal velocity and, 
therefore, the absorption of the helicon wave in the collisionless plasma by electrons due to the 
electron Landau damping is a negligibly weak effect. That is why the experimentally observed
\cite{Boswell} unusually high absorption rate of the helicon wave, that testified about strong 
interaction of the helicon wave with electrons, was unpredictable. 
Although a large number of studies have been curried, the mystery of why helicon discharges 
are so efficient is still unresolved. 

The anomalous absorption of helicons and plasma heating was observed a very long time ago in 
the first experimental studies\cite{Grigor'eva, Porkolab} of the basic plasma physics processes. 
It was claimed in these papers that the anomalous absorption of a large amplitude whistler wave
was caused by the development of the current driven ion acoustic instability\cite{Grigor'eva} or by 
the development of the resonant decay instability\cite{Porkolab}. The Boswell's experiments 
gave impetus to the active theoretical investigations of the plasma instabilities driven by the 
helicon wave. It was found that the development of the 
parametric kinetic\cite{Akhiezer} and decay\cite{Aliev} ion acoustic instabilities, 
originated from the oscillatory motion of electrons relative to 
ions in the pumping helicon field, may be the cause of the anomalous absorption of the helicon wave 
and of the anomalous heating of electrons, resulted from the interaction of electrons with 
ion acoustic turbulence. Since then, the plasma turbulence in helicon plasma 
was investigated in several experiments \cite{Altukhov, Kramer}, in which the 
detected short scale fluctuations were identified as the ion acoustic  waves 
and was found that the level of these fluctuations increases with RF power. 

The developed theory of the parametric instabilities of the helicon plasma is grounded on the 
approximation of the spatially uniform helicon wave. However, the helicon wave field in cylindrical 
helicon sources is as a rule spatially inhomogeneous with radial inhomogeneity length 
comparable with, or less than, the radius of plasma cylinder. The effect of the cylindrical 
geometry of the plasma and of the helicon wave field, and the effect of the spatial 
inhomogeneity of the helicon wave on the parametric microturbulence is usually ignored assuming 
that the approximation of the uniform helicon field is sufficient for the proper description 
of the parametric instabilities the wavelengths of which are much less than the radial 
inhomogeneity scale length of the helicon 
wave in plasma cylinder. It was found in Refs.\cite{Mikhailenko, Mikhailenko1}, however, 
that the electromagnetic field inhomogeneity in the inductive plasma sources may be the 
powerful source of the instabilities development. It was derived\cite{Mikhailenko, Mikhailenko1} 
that the accelerated motion of electrons relative to ions under the action of the 
ponderomotive force, formed in the skin layer of the inductively coupled plasma, 
is the much more stronger source of the 
instabilities development than the quiver motion of the electron in the electromagnetic field.
The spatial structure of the helicon wave in the helicon sources depends on the 
antenna design, on the magnitude of the confined magnetic field, on the input RF power and RF 
frequency, on the magnitude and radial profile of the electron density, ets. 
The focus of this paper is the two-scale kinetic theory of the 
microscale instabilities of the helicon plasma driven by the macroscale plasma flow, 
formed by the radially inhomogeneous helicon wave in the radially inhomogeneous plasma. 
In Sec. \ref{sec2}, using the Vlasov-Poisson 
model we develop the theory of the stability of the cylindrical plasma in the field of 
the azimuthally symmetric $(m=0)$ radially inhomogeneous helicon wave. In high frequency  helicon 
wave field, electrons experience the oscillating motion relative to the 
practically unmovable ions. This motion is known as a source of the development of the parametric 
instabilities\cite{Akhiezer, Aliev}, investigated usually employing the approximation 
of the uniform helicon wave. 
The developed in this paper theory reveals new macroscale effect of the azimuthal 
steady rotation of electrons with radially inhomogeneous angular velocity. This effect was 
detected experimentally in the helicon plasma source, that uses azimuthally symmetric antenna,
long time ago\cite{Tynan}, but was not explained yet. We found that this rotation is caused by the 
radial inhomogeneity of the cylindrical helicon wave and is spatially localized 
in the region of strong gradients in the helicon wave. The main result of this section is the 
derived basic integral equation for the electrostatic potential, which governs the 
microinstabilities of the radially inhomogeneous cylindrical plasma driven by the azimuthally 
symmetric inhomogeneous helicon wave. The solution of this integral equation for the short scale 
high frequency ion acoustic instability is presented in Sec. \ref{sec3}. The Conclusions are 
given in Sec. \ref{sec4}.

\section{Basic equations}\label{sec2} 

We consider an axially symmetric plasma in a uniform axial magnetic field $\mathbf{B}_{0}$, 
directed along $z$ axes, and in the electric 
$\mathbf{E}_{1}=\mathbf{E}_{1r}\left(r, z, t \right)+
\mathbf{E}_{1\varphi}\left(r, z, t \right)$, 
and magnetic $\mathbf{B}_{1z}\left(r,z,t \right)$ fields  of the azimuthally symmetric helicon wave, 
excited by the loop antenna located 
on the boundary $r=r_{0}$ of the cylindrical chamber. The helicon wave field is given by the 
relations
\begin{eqnarray}
&\displaystyle
E_{1r}\left(r,z,t \right)=E_{1r}\left(r \right) \sin\left(k_{0z}z-\omega_{0}t \right) ,
\nonumber
\\  
&\displaystyle
E_{1\varphi}\left(r,z,t \right)=E_{1\varphi}\left(r \right) \cos\left(k_{0z}z-\omega_{0}t \right),
\nonumber
\\  
&\displaystyle 
B_{1z}\left(r,z,t \right)=B_{1}\left(r \right)  \sin\left(k_{0z}z-\omega_{0}t \right),
\label{1}
\end{eqnarray}
where $\omega_{0}$ is the frequency of the helicon wave, $k_{0z}$ is the wavenumber component along 
magnetic field $B_{0}$. In such a fields, a plasma 
in equilibrium has an azimuthally symmetric radially inhomogeneous density profile. The radial 
profiles of the electric and magnetic fields
depends on the plasma density profiles and differ from the fields 
\begin{eqnarray}
&\displaystyle
E_{1r}\left(r \right)=E_{1r}J_{1}\left(k_{0\perp}r \right), 
\nonumber
\\  
&\displaystyle
E_{1\varphi}\left(r\right)=E_{1\varphi}J_{1}\left(k_{0\perp}r \right), 
\nonumber
\\  
&\displaystyle
B_{1z}\left(r\right)=B_{1}J_{0}\left(k_{0\perp}r \right),
\label{2}
\end{eqnarray}
where $k_{0\perp}=\omega_{0}\omega_{pe}^{2}/\left(\omega_{ce}k_{0z}c^{2}\right)$, $J_{0,1}
\left(k_{0\perp}r\right) $ are the  Bessel functions, known for the helicon plasmas with a uniform 
density\cite{Chen}. In the cylindrical coordinates $r$, $\varphi$, $z$ for the electron position and 
$v_{\bot}$, $\phi$, $v_{z}$ for the electron velocity, the Vlasov equation for electrons has a form

\begin{eqnarray}
&\displaystyle
\frac{\partial F_{e}}{\partial t}+v_{\perp}\cos\phi\frac{\partial F_{e}}{\partial r}
+\frac{v_{\perp}}{r}\sin\phi\frac{\partial F_{e}}{\partial \varphi}
+v_{z}\frac{\partial F_{e}}{\partial z}
\nonumber
\\  
&\displaystyle
+\frac{e}{m_{e}}\left(\sin\phi E_{\varphi}+\cos\phi E_{r} \right) \frac{\partial F_{e}}{\partial 
v_{\bot}}
\nonumber
\\  
&\displaystyle
-\left[\omega_{ce}+ \frac{v_{\perp}}{r}\sin\phi+\frac{e}{m_{e}v_{\bot}}
+\frac{eB_{1z}\left(r,z,t \right) }{m_{e}c}\right. 
\nonumber
\\  
&\displaystyle
\left(\sin\phi E_{r}-\cos\phi E_{\varphi}\right)\left]\frac{\partial F_{e}}{\partial \phi}\right. 
+\frac{e}{m_{e}}E_{z}\frac{\partial F_{e}}{\partial v_{z}}=0,
\label{3}
\end{eqnarray}
where $\omega_{ce}=eB_{0}/m_{e}c$ is the electron cyclotron frequency. In this equation, 
electric field $\mathbf{E}\left(r,\varphi,z,t 
\right)=\mathbf{E}_{r}\left(r,\varphi,z,t \right) +\mathbf{E}_{\varphi}\left(r,
\varphi,z,t \right)+ \mathbf{E}_{z}\left(r,\varphi,z,t \right)$ is
\begin{eqnarray}
&\displaystyle
\mathbf{E}\left(r,\varphi,z,t \right)=\mathbf{E}_{1}\left(r,z,t \right)
+\mathbf{\tilde{E}}\left(r,\varphi,z,t \right),
\label{4}
\end{eqnarray}
where 
\begin{eqnarray}
&\displaystyle
\mathbf{\tilde{E}}\left(r,\varphi,z,t \right)=-\nabla \Phi \left(r,\varphi,z,t \right)
\label{5}
\end{eqnarray}
is the electrostatic electric field of the plasma response on the helicon wave. 
This self consistent electric field is determined by the Poisson equation, 
\begin{eqnarray}
&\displaystyle 
-\bigtriangleup\Phi \left(r,\varphi,z,t \right)=
4\pi\sum_{\alpha=i,e} e_{\alpha}\int f_{\alpha}\left(\mathbf{v}, r,\varphi,z, t \right)d{\bf v}, 
\label{6}
\end{eqnarray}
in which $f_{\alpha}$ is the fluctuating part of the distribution function 
$F_{\alpha}$, $f_{\alpha} =F_{\alpha}-F_{0\alpha}$, 
where $F_{0\alpha}$ is the equilibrium distribution function.
The Vlasov equation (\ref{3}) for electron and ion components, and the Poisson equation (\ref{6}) 
compose the basic system of equations of our studies. 

\section{The solution of the Vlasov equation for the 
cylindrical plasma in the field of the azimuthally symmetric helicon wave}\label{sec3} 

\begin{figure}[ht]
\includegraphics[width=0.4\textwidth]{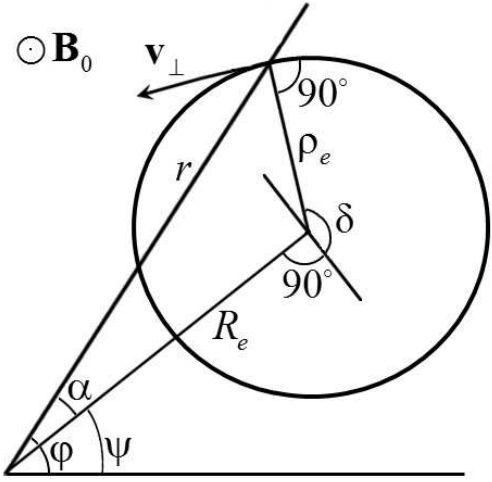}
\caption{\label{fig1} The geometric interpretation of the cylindrical guiding center 
coordinates for an electron.}
\end{figure}

The Vlasov equation for the electron distribution function $F_{e}\left(v_{\bot},\phi,v_{z},r,
\varphi,z,t \right)$ of the helicon sources contains two 
different spatial scales: the macroscale of the  radial inhomogeneity of the helicon wave, 
which is commensurable with radial scale of the plasma density inhomogeneity, and the microscale 
commensurable with the thermal Larmor radius of electrons. In this section, we derive the two-scale 
solution of Eq. (\ref{3}) by the transformation of the $r,\varphi, v_{\bot},\phi$ coordinates 
to the cylindrical guiding center 
coordinates $R_{e}, \psi, \rho_{e}, \delta$ determined by the relations\cite{Chibisov} 
\begin{eqnarray} 
&\displaystyle
R_{e}^{2}=\frac{1}{\omega_{ce}^2}\left(v^{2}_{\perp}+2v_{\perp}r\omega_{ce}\sin\phi+r^{2}
\omega_{ce}^{2} \right), 
\label{7}
\end{eqnarray}
\begin{eqnarray}
&\displaystyle
\psi=\varphi-\alpha, 
\label{8}
\end{eqnarray}
\begin{eqnarray}
&\displaystyle
\rho^{2}_{e}=\frac{v_{\perp}^2}{\omega_{ce}^{2}},
\label{9}
\end{eqnarray}
\begin{eqnarray}
&\displaystyle
\delta=\phi+\alpha,
\label{10}
\end{eqnarray}
\begin{eqnarray*}
&\displaystyle
\alpha=\arcsin\left[\frac{\cos\phi}{
\left(1+v^{-2}_{\perp}\left(r^{2}\omega_{ce}^{2}+2v_{\perp}r\omega_{ce}
\sin\phi \right)\right)^{1/2}}\right].
\end{eqnarray*}
The geometric interpretation of the cylindrical guiding center coordinates for an electron is 
presented in Fig. \ref{fig1}.  
In  coordinates $R_{e}$, $\psi$, $\rho_{e}$, $\delta$, $z$, $t$, Eq. (\ref{3}) 
transforms to the following equation for $F_{e}\left(R_{e},\psi, \rho_{e}, \delta, z, t\right)$:
\begin{eqnarray}
&\displaystyle
\frac{\partial F_{e}}{\partial t}+v_{z}\frac{\partial F_{e}}{\partial z}-\omega_{ce}\frac{\partial 
F_{e}}{\partial \delta}
\nonumber
\\  
&\displaystyle
+\frac{c}{B_{0}}\frac{1}{\left(R^{2}_{e}-2\rho_{e}R_{e}\sin \delta + \rho^{2}_{e}\right)^{1/2}}
\nonumber
\\  
&\displaystyle
\times
\left[\left(E_{1r}\rho_{e}\cos \delta  +
E_{1\varphi}\left(R_{e}-\rho_{e}\sin \delta\right)\right)\frac{\partial F_{e}}{\partial R_{e}}\right.
\nonumber
\\  
&\displaystyle
-\left.\left(\left(1-\frac{\rho_{e}}{R_{e}}\sin\delta\right)E_{1r}+\frac{\rho_{e}}{R_{e}}\cos \delta 
E_{1\varphi}\right)\frac{\partial F_{e}}{\partial\psi}\right.
\nonumber
\\  
&\displaystyle
\left.+\left(E_{1r}R_{e}\cos \delta +E_{1\varphi}\left(R_{e}\sin\delta -\rho_{e}\right)\right)
\frac{\partial F_{e}}{\partial \rho_{e}}
\right.
\nonumber
\\  
&\displaystyle
+\left(E_{1r}\left(2-\sin \delta \frac{\left(R^{2}_{e}+\rho^{2}_{e}\right)}{R_{e}
\rho_{e}}\right) \right. 
\nonumber
\\  
&\displaystyle
\left. \left. 
+E_{1\varphi}\cos \delta\frac{\left(R^{2}_{e}+\rho^{2}_{e}\right)}{R_{e}\rho_{e}}\right)
\frac{\partial 
F_{e}}{\partial\delta}\right]
\nonumber
\\  
&\displaystyle
-\frac{e}{m_{e}c}B_{1z}\left[\rho_{e}\cos \delta\frac{\partial F_{e}}{\partial R_{e}}
+ \frac{\partial F_{e}}{\partial \delta}\right. 
\nonumber
\\  
&\displaystyle
\left. 
-\sin \delta \frac{\rho_{e}}
{R_{e}}\left(\frac{\partial F_{e}}{\partial \delta}
-\frac{\partial F_{e}}{\partial \psi}\right)\right]
\nonumber
\\  
&\displaystyle
-\frac{c}{B_{0}R_{e}}\left(\frac{\partial \Phi}{\partial \psi}-\frac{\partial \Phi}{\partial \delta}
\right)\frac{\partial F_{e}}{\partial R_{e}}
\nonumber
\\  
&\displaystyle
+\frac{c}{B_{0}\rho_{e}}\frac{\partial \Phi}{\partial 
\delta}\frac{\partial F_{e}}{\partial\rho_{e}}
-\frac{e}{m_{e}}\frac{\partial \Phi}{\partial z}\frac{\partial F_{e}}{\partial v_{ez}}=0.
\label{11}
\end{eqnarray}
In the helicon discharge,  $\rho_{e}\ll R_{e}$, excluding small region $R_{e}\sim \rho_{e}$ of the 
discharge center. Equation (\ref{11}), 
in which the terms on the order of $\rho_{e}/R_{e}\ll 1$ are omitted, becomes
\begin{eqnarray}
&\displaystyle
\frac{\partial F_{e}}{\partial t}+v_{z}\frac{\partial F_{e}}{\partial z}
+\frac{c}{B_{0}}E_{1\varphi}\frac{\partial F_{e}}{\partial R_{e}} - \frac{c}{B_{0}}E_{1r}
\frac{\partial F_{e}}{R_{e}\partial\psi}
\nonumber
\\  
&\displaystyle
+\frac{c}{B_{0}}\left(E_{1r}\cos \delta+E_{1\varphi}\sin \delta\right)\frac{\partial F_{e}}
{\partial \rho_{e}}
\nonumber
\\  
&\displaystyle
+\left(2\frac{c}{B_{0}R_{e}}E_{1r}-\omega_{ce}\right. 
\nonumber
\\  
&\displaystyle
\left. 
-\frac{c}{B_{0}\rho_{e}}\frac{1}{\rho_{e}}
\left(E_{1r}\sin \delta-E_{1\varphi}\cos \delta\right)\right)\frac{\partial F_{e}}{\partial\delta}
\nonumber
\\  
&\displaystyle
-\frac{e}{m_{e}c}B_{1z}\frac{\partial F_{e}}{\rho_{e}\partial\delta}
-\frac{c}{B_{0}R_{e}}\left(\frac{\partial \Phi}{\partial \psi}-\frac{\partial \Phi}{\partial \delta}
\right)\frac{\partial 
F_{e}}{\partial R_{e}}
\nonumber
\\  
&\displaystyle
+\frac{c}{B_{0}\rho_{e}}\frac{\partial \Phi}{\partial \delta}\frac{\partial 
F_{e}}{\partial\rho_{e}}
-\frac{e}{m_{e}}\frac{\partial \Phi}{\partial z}\frac{\partial F_{e}}{\partial v_{ez}}=0.
\label{12}
\end{eqnarray}
The Vlasov equation (\ref{12}) with $\Phi=0$ is the equation for the equilibrium 
electron distribution function $F_{e0}$. Consider now the system of equations for the 
characteristics of equation for $F_{e0}$,
\begin{eqnarray}
&\displaystyle
dt=\frac{dR_{e}}{\frac{c}{B_{0}}E_{1\varphi}}=\frac{d\psi}{-\frac{c}{B_{0}R_{e}}E_{1r}}=\frac{d
\rho_{e}}{\frac{c}
{B_{0}}\left(E_{1r}\cos \delta+E_{1\varphi}\sin\delta\right)}
\nonumber
\\  
&\displaystyle
=d\delta\left[ -\omega_{ce}+2\frac{c}{B_{0}R_{e}}E_{1r}-\frac{eB_{1z}}{cm_{e}}\right. 
\nonumber
\\  
&\displaystyle
\left. 
-\frac{c}{B_{0}\rho_{e}}\left(E_{1r}\sin 
\delta-E_{1\varphi} \cos \delta\right)\right] ^{-1}=\frac{dz}{v_{z}}.
\label{13}
\end{eqnarray}
With approximations $E_{1r}\left(r\right)\approx E_{1r}\left(R_{e}\right)$ and 
$E_{1\varphi}\left(r\right)\approx E_{1\varphi}\left(R_{e}\right)$, which follows from the relation
\begin{eqnarray}
&\displaystyle
r=\left(R^{2}_{e}-2\rho_{e}R_{e}\sin\delta+\rho^{2}_{e} \right)^{1/2}\approx R_{e}
\label{14}
\end{eqnarray} 
in the limit $\rho_{e}\ll R_{e}$, the system of equations for the guiding center coordinates 
$R_{e}$, $\psi$
\begin{eqnarray}
&\displaystyle
dt=\frac{dR_{e}}{\frac{c}{B_{0}}E_{1\varphi}\left(R_{e}\right)\cos \left(\omega_{0}t
-k_{0z}\left(z_{1}+v_{z}t\right)\right)}
\nonumber
\\  
&\displaystyle
=\frac{d\psi}{-\frac{c}{B_{0}R_{e}}E_{1r}\left(R_{e}\right)\sin \left(\omega_{0}t-k_{0z}\left(z_{1}
+v_{z}t\right)\right)},
\label{15}
\end{eqnarray}
where $z_{1}=z-v_{z}t$ is the integral of system (\ref{13}), becomes separate from the system 
of equation for the coordinates $\rho_{e}$ and $\delta$ of the Larmor motion. 

Now we consider the approximate solution to the nonlinear equation for $R_{e}$,
\begin{eqnarray}
&\displaystyle
\frac{dR_{e}}{dt}=E_{1\varphi}\left(R_{e}\right)\cos\left(\omega_{0}t-k_{z0}z_{1}\right), 
\label{16}
\end{eqnarray}
in which the term $k_{0z}v_{z}t$ is omitted, because for the helicon wave $\omega_{0}\gg k_{0z}v_{Te}$, where $v_{Te}$ 
is the electron thermal velocity.  That solution is derived in the form
\begin{eqnarray}
&\displaystyle
R_{e}\left(t \right) \approx R_{e1}+\frac{c}{B_{0}\omega_{0}}E_{1\varphi}\left(R_{e1} \right)
\sin\left( \omega_{0}t-k_{z0}z_{1}\right) 
\nonumber
\\  
&\displaystyle
-\frac{c^{2}}{B_{0}^{2}\omega_{0}^{2}}\frac{1}{4}\sin^{2}\left( \omega_{0}t-k_{z0}z_{1}\right) 
\frac{d}{dR_{e1}}E_{1\varphi}^{2}\left(R_{e1} \right),
\label{17}
\end{eqnarray}
which is the  power series expansion in $\left|\xi/R_{e1}\right|\ll 1$, where $\xi=\frac{c}{B_{0}
\omega_{0}}E_{1\varphi}\left(R_{e1} \right)$ 
is the amplitude of the displacement of an electron along the coordinate $R_{e}$ and $R_{e1}$ 
is the integral of Eq. (\ref{16}).

The approximate solution to the equation for the angle $\psi$,
\begin{eqnarray}
&\displaystyle
\frac{d\psi}{dt}=\frac{c}{B_{0}R_{e}}E_{1r}\left(R_{e} \right) \sin\left( \omega_{0}t-k_{z0}z_{1}
\right) 
\label{18}
\end{eqnarray}
in the form of the power series expansion in $\left|\xi/R_{e1}\right|\ll 1$ with accounting 
for the terms on the first and second order of $\left|\xi/R_{e1}\right|^{2}$ has a form
\begin{eqnarray}
&\displaystyle
\psi\approx \psi_{1}-\frac{c}{B_{0}R_{e1}\omega_{0}}E_{1r}\left(R_{e1} \right) 
\cos\left( \omega_{0}t-k_{z0}z_{1}\right)
\nonumber
\\  
&\displaystyle
-\frac{c^{2}}{2B_{0}^{2}R_{e1}^{2}\omega_{0}}E_{1\varphi}\left(R_{e1} \right)
\left(E_{1r}\left(R_{e1} \right)-R_{e1}\frac{\partial E_{1r}\left(R_{e1} \right)}{\partial R_{e1}}
\right) 
\nonumber
\\  
&\displaystyle
\times
\left[ t-\frac{1}{2\omega_{0}}\sin2\left( \omega_{0}t-k_{z0}z_{1}\right)
\right], 
\label{19}
\end{eqnarray}
It contains the terms oscillating on frequencies $\omega_{0}$ and $2\omega_{0}$, and the term 
corresponding to the rotation of the guiding 
center coordinate with stationary radially inhomogeneous angular velocity $\Omega_{e}\left(R_{e1}
\right)$,
\begin{eqnarray}
&\displaystyle
\Omega_{e}\left(R_{e1}\right)=  \frac{c^{2}E_{1\varphi}\left(R_{e1}
\right)}{2B_{0}^{2}R_{e1}^{2}\omega_{0}}
\nonumber
\\  
&\displaystyle
\times
\left[ E_{1r}\left(R_{e1} \right)
-R_{e1}\frac{\partial E_{1r}\left(R_{e1} \right)}{\partial R_{e1}}
\right]. 
\label{20}
\end{eqnarray}
It follows from (\ref{19}) that the secular growth of $\psi$ becomes dominant at time $t$, at which 
\begin{eqnarray}
&\displaystyle
\omega_{0}t>2\frac{B_{0}R_{e1}\omega_{0}}{cE_{1\varphi}}>1.
\label{21}
\end{eqnarray}
Equation (\ref{21}) determines the condition under which the theory of the parametric instabilities, 
based on the approximation of the uniform helicon wave, becomes wrong.  In what follows, we assume 
that condition (\ref{21}) is valid for the selected plasma and helicon wave parameters and the sources 
of the instabilities development will be different from the oscillating motion of electrons 
relative to ions. For time $t$, for which condition (\ref{21}) is valid, the approximation 
\begin{eqnarray}
&\displaystyle
\psi=\psi_{1}-\Omega_{e}\left(R_{e1}\right)t
\label{22}
\end{eqnarray}
may be used. For $B_{0}=100$G, $\omega_{0}=10^{7}$c$^{-1}$, $R_{e1}=2.5$ cm, $E_{1\varphi}=5$V/cm, 
condition (\ref{21}) is satisfied for $\omega_{0}t>10$, i. e. at time $t> 10^{-6}$ c.

Because $\omega_{ce}$ is much larger than any other term in the equation for $d\delta/dt$ 
of system (\ref{13}), the solution for $\delta_{1}$ with a great accuracy is
\begin{eqnarray}
&\displaystyle
\delta=\delta_{1}-\omega_{ce}t.
\label{23}
\end{eqnarray}
The solution of the equation for the radius $\rho_{e}$ of the electron Larmor orbit,
\begin{eqnarray}
&\displaystyle
\frac{d\rho_{e}}{dt}=\frac{c}{B_{0}}\Big[
E_{1r}\left(R_{e} \right)
\sin \left(k_{z0}z_{1}-\omega_{0}t\right)\cos\left(\delta_{1}-\omega_{ce}t \right) 
\nonumber
\\  
&\displaystyle
+E_{1\varphi}\left(R_{e} \right)
\cos \left(\omega_{0}t-k_{z0}z_{1}\right)\sin\left(\delta_{1}-\omega_{ce}t \right) 
\Big],
\label{24}
\end{eqnarray}
where $\delta_{1}$ is given by Eq. (\ref{23}) is
\begin{eqnarray}
&\displaystyle
\rho_{e}=\rho_{e1}-\frac{c}{B_{0}\omega_{ce}}
\Big[E_{1r}\left(R_{e1} \right)
\sin \left(\omega_{0}t-k_{z0}z_{1}\right)
\nonumber
\\  
&\displaystyle
\times\sin\left(\omega_{ce}t-\delta_{1} \right) 
\nonumber
\\  
&\displaystyle
-E_{1\varphi}\left(R_{e1} \right)
\cos \left(\omega_{0}t-k_{z0}z_{1}\right)\cos\left(\omega_{ce}t-\delta_{1} \right) 
\Big]. 
\label{25}
\end{eqnarray}
with accuracy to terms of the order of $O\left(\frac{\omega_{0}}{\omega_{ce}}\ll 1 \right)$. 
It is easy to check that the Vlasov equation for the equilibrium electron distribution function 
$F_{e0}$ in variables $R_{e1}, \psi_{1}, \rho_{e1}, \delta_{1}, v_{z}, z_{1}$ determined by the 
solutions (\ref{17}), (\ref{19}), (\ref{24}), (\ref{25}), reduces to the equation 
\begin{eqnarray*}
&\displaystyle
\frac{\partial F_{e0}}{\partial t}=0,
\end{eqnarray*}
and, therefore, $F_{e0}= F_{e0}\left(R_{e1}, \psi_{1}, \rho_{e1}, \delta_{1}, v_{z}, z_{1}\right)$, 
and does not depend on time variable.

The equation for the perturbation\\ 
$f_{e}\left(R_{e1}, \psi_{1}, \rho_{e1}, \delta_{1}, v_{z}, z_{1}, t\right)$ becomes
\begin{eqnarray}
&\displaystyle
\frac{\partial}{\partial t}f_{e}\left(R_{e1}, \psi_{1}, \rho_{e1}, 
\delta_{1}, v_{z}, z_{1}, t\right)
\nonumber
\\  
&\displaystyle
=\frac{c}{B_{0}R_{e}}\left(\frac{\partial \Phi}{\partial \psi}
- \frac{\partial\Phi}{\partial \delta}
\right)	\frac{\partial F_{e0}\left(R_{e1}, \rho_{e1}, v_{z} \right)}{\partial R_{e1}}
\nonumber
\\  
&\displaystyle
-\frac{c}{B_{0}}\frac{1}{\rho_{e}}\frac{\partial\Phi}{\partial \delta}
\frac{\partial F_{e0}}{\partial \rho_{e1}} 
+\frac{e}{m_{e}}\frac{\partial\Phi}{\partial z_{1}}
\frac{\partial F_{e0}}{\partial v_{ez}}. 
\label{26}
\end{eqnarray}
In this equation, potential $\Phi\left(R_{e1}, \psi_{1}, \rho_{e1},\delta_{1}, v_{z},z_{1},t\right)$ 
is presented in the form of the Fourier-Bessel transformation,
\begin{eqnarray}
&\displaystyle
\Phi\left(R_{e1}, \psi_{1}, \rho_{e1},\delta_{1}, v_{z},z_{1},t\right)
=\sum\limits^{\infty}_{m=-\infty}\sum\limits^{\infty}_{n=-\infty}
\nonumber
\\  
&\displaystyle
\times
\int
dk_{\perp}k_{\perp}dk_{z}d\theta d\omega \Phi\left(k_{\perp}, \theta, k_{z}, \omega \right)
\nonumber
\\  
&\displaystyle
\times
J_{n}\left(k_{\perp}\rho_{e1} \right) J_{n+m}\left(k_{\perp}R_{e1} \right) 
\nonumber
\\  
&\displaystyle
\times
\exp\left[-in\left(\delta_{1}-\omega_{ce}t\right)-im\left(\theta-\psi_{1}
+\Omega_{e}\left(R_{e1}\right)t \right)\right.
\nonumber
\\  
&\displaystyle
\left.+i\left(m+n \right) \frac{\pi}{2}-i\left(\omega-k_{z}v_{z}\right)t+ik_{z}z_{1} \right], 
\label{27}
\end{eqnarray}
It was derived from the Fourier-Bessel transform 
\begin{eqnarray}
&\displaystyle
\Phi\left(r,\varphi,z,t \right) =\int \Phi\left(\mathbf{k},\omega \right) e^{-\omega t}
d\omega 
\nonumber
\\  
&\displaystyle
\times
e^{ik_{\perp}r\cos\left(\theta-\varphi \right) }k_{\perp}dk_{\perp}
d\theta e^{ik_{z}z}dk_{z},
\label{28}
\end{eqnarray}
in which the identity
\begin{eqnarray}
&\displaystyle
k_{\perp}r\cos\left(\theta-\varphi \right) =k_{\perp}R_{e}\cos\left(\theta-\psi \right) 
\nonumber
\\  
&\displaystyle
+k_{\perp}\rho_{e}\sin\left(\theta-\psi-\delta \right) 
\label{29}
\end{eqnarray}
and Eqs. (\ref{22}) and (\ref{23}) were employed.

\section{The ion acoustic instability of the cylindrical plasma in the field of the azimuthally 
symmetric helicon wave}\label{sec4} 
By using the solution for $f_{e}$ of Eq. (\ref{26}) in the Poisson equation (\ref{6}), we 
derive the Fourier transformation of the Poisson equation,
\begin{eqnarray}
k^{2}\Phi\left(\mathbf{k},\omega \right)= 4\pi e_{i}n_{i}\left(\mathbf{k},\omega \right)+
4\pi e n_{e}\left(\mathbf{k},\omega \right),
\label{30}
\end{eqnarray}
where
\begin{eqnarray}
&\displaystyle
n_{e}\left(\mathbf{k},\omega\right)=\sum^{\infty}_{m=-\infty}e^{-im\theta}n_{em}
\left(k_{\bot}, k_{z}, \omega\right)
\nonumber
\\  
&\displaystyle
=-\omega^{2}_{ce}\frac{e}{m_{e}}\sum^{\infty}_{m=-\infty}e^{-im\theta}\sum^{\infty}_{n=-\infty}\int
\limits^{\infty}_{0}dR_{e1}R_{e1}\int \limits^{\infty}_{-\infty}dv_{z}
\nonumber
\\  
&\displaystyle
\times
\int\limits^{\infty}_{0}d\rho_{e1}\rho_{e1}
\int\limits_{0}^{\infty}dk_{1\bot}k_{1\bot}\Phi_{m}\left(k_{1\bot},k_{z},\omega\right)J_{n}\left(k_{\bot}\rho_{e1}\right)
\nonumber
\\  
&\displaystyle
\times\frac{J_{n}\left(k_{1\bot}\rho_{e1}\right)J_{n+m}
\left(k_{\bot}R_{e1}\right)J_{n+m}\left(k_{1\bot}R_{e1}\right)}
{\omega-n\omega_{ce}+m\Omega_{e}\left(R_{e1}\right)-k_{z}v_{z}}
\nonumber
\\  
&\displaystyle
\times \left[\frac{m+n}{\omega_{ce}R_{e1}}\frac{\partial F_{e0}}{\partial 
R_{e1}}+\frac{n}{\omega_{ce}}\frac{1}{\rho_{e1}}\frac{\partial F_{e0}}{\partial \rho_{e1}}+k_{z}
\frac{\partial F_{e0}}{\partial v_{z}}\right]. 
\label{31}
\end{eqnarray}
In Eq. (\ref{31}), the $m$-th Fourier harmonic $\Phi_{m}\left(k_{\bot}, k_{z}, \omega\right)$ of the 
potential $\Phi$ is determined by the relation
\begin{eqnarray}
&\displaystyle
\Phi_{m}\left(k_{\bot},k_{z},\omega\right) =\frac{1}{2\pi}\int d\theta_{1}\Phi\left(k_{\bot},
\theta_{1},k_{z},\omega \right)e^{im\theta_{1}}.
\label{32}
\end{eqnarray}

The equation for $\Phi_{m}\left(k_{\bot},k_{z},\omega\right)$ follows from (\ref{30})--(\ref{32}),
\begin{eqnarray}
&\displaystyle
\Phi_{m}\left(k_{\bot},k_{z},\omega\right)\left(1-\frac{\omega^{2}_{pi}}{\omega^{2}}\right)
\nonumber
\\  
&\displaystyle
+8\pi^{2}\frac{e^{2}}{k^{2}m_{e}}\omega^{2}_{ce}
\sum^{\infty}_{n=-\infty}\int\limits^{\infty}_{0}
dR_{e1}R_{e1}\int \limits^{\infty}_{-\infty}dv_{z}
\nonumber
\\  
&\displaystyle
\times
\int\limits^{\infty}_{0}d\rho_{e1}\rho_{e1}
\int\limits_{0}^{\infty}dk_{1\bot}k_{1\bot}\Phi_{m}\left(k_{1\bot},k_{z},\omega\right)
J_{n}\left(k_{\bot}\rho_{e1}\right)
\nonumber
\\  
&\displaystyle
\times\frac{J_{n}\left(k_{1\bot}\rho_{e1}\right)J_{n+m}
\left(k_{\bot}R_{e1}\right)J_{n+m}\left(k_{1\bot}R_{e1}\right)}
{\omega-n\omega_{ce}+m\Omega_{e}\left(R_{e1}\right)-k_{z}v_{z}}
\nonumber
\\  
&\displaystyle
\times \left[\frac{m+n}{\omega_{ce}R_{e1}}\frac{\partial F_{e0}}{\partial 
R_{e1}}+\frac{n}{\omega_{ce}\rho_{e1}}\frac{\partial F_{e0}}{\partial \rho_{e1}}+k_{z}
\frac{\partial F_{e0}}{\partial v_{z}}\right]=0.
\label{33}
\end{eqnarray}
In Eq. (\ref{33}), the approximation of the unmovable ions in the helicon wave was used, 
which is applicable for the treating of the instabilities with
the growth rate $\gamma\left(\mathbf{k}\right)\gg \omega_{ci}/2\pi$ and $k_{\bot}\rho_{i}\gg 1$. 
Here, we derive the solution to Eq. (\ref{33}) in the short wavelength limit
\begin{eqnarray}
&\displaystyle
k_{\bot}R_{e1}\sim m\gg 1, 
\label{34}
\end{eqnarray}
analysed in Ref. \cite{Mikhailenko2} in the studies of the drift turbulence of the 
azimuthally symmetric radially nonuniform plasma and in Ref.\cite{Mikhailenko3} in the studies 
of the shear flow driven ion cyclotron and ion acoustic instabilities of the cylindrical 
inhomogeneous plasma. It follows from Refs.\cite{Mikhailenko2,Mikhailenko3} that the 
solution to Eq. (\ref{33}) under condition (\ref{34}), is 
\begin{eqnarray}
&\displaystyle
\Phi_{m}\left(k_{\bot},k_{z},\omega\right)= \Phi_{m}\left(k_{\bot},k_{z},\omega_{m}
\left(k_{\bot},k_{z}\right)\right) 
\label{35}
\end{eqnarray}
for $ \omega=\omega_{m}\left(k_{\bot},k_{z}\right)$ and $\Phi_{m}\left(k_{\bot},k_{z},\omega\right) 
=0$ for $ \omega\neq\omega_{m}\left(k_{\bot},k_{z}\right)$, 
where $\omega_{m}\left(k_{\bot},k_{z}\right)$ is the solution to the equation
\begin{eqnarray}
&\displaystyle
\varepsilon_{m}\left(k_{\bot},k_{z}, \omega\right)=1-\frac{\omega^{2}_{pi}}{\omega^{2}}
\nonumber
\\  
&\displaystyle
+8\pi^{2}\frac{e^{2}}{k^{2}m_{e}}\omega^{2}_{ce}
\sum^{\infty}_{n=-\infty}\int\limits^{\infty}_{0}d\rho_{e1}\rho_{e1}
\int \limits^{\infty}_{-\infty}dv_{z}
\nonumber
\\  
&\displaystyle
\times\frac{J^{2}_{n}\left(k_{\bot}\rho_{e1}\right)}
{\left(\omega-n\omega_{ce}+m\Omega_{e}\left(R_{e1}\right)-k_{z}v_{z}\right)}
\left[\frac{m+n}{\omega_{ce}R_{e1}}\frac{\partial F_{e0}}{\partial 
R_{e1}}\right. 
\nonumber
\\  
&\displaystyle
\left. 
+\frac{n}{\omega_{ce}}\frac{1}{\rho_{e1}}\frac{\partial F_{e0}}{\partial \rho_{e1}}+k_{z}
\frac{\partial F_{e0}}{\partial v_{z}}\right]_{R_{e1}=R_{e0}=\frac{m}{k_{\bot}}}=0.
\label{36}
\end{eqnarray}
The cylindrical plasma excited by the cylindrically 
symmetric helicon wave has an azimuthally symmetric radially inhomogeneous density 
profile. For the Maxwellian distribution $F_{e0}\left(\rho_{e}, v_{z}, R_{e0}\right)$, 
\begin{eqnarray}
&\displaystyle
F_{e0}\left(\rho_{e}, v_{z}, R_{e0}\right)=\frac{n_{e0}\left(R_{e0}\right)}{\left(2\pi\right)^{3/2}
v^{3}_{Te}}
\nonumber
\\  
&\displaystyle
\times
\exp\left[ 
-\frac{\rho^{2}_{e1}}{2\rho^{2}_{Te}}-\frac{v^{2}_{z}}{v^{2}_{Te}}\right] , 
\label{37}
\end{eqnarray}
where $\rho_{Te}=v_{Te}\left(R_{e0}\right)/\omega_{ce}$, and $v^{2}_{Te}\left(R_{e0}\right)
=T_{e}\left(R_{e0}\right)/m_{e}$, we derive from Eq. (\ref{36}) the following dispersion 
equation for the low frequency perturbations with $\omega \ll \omega_{ce}$:
\begin{eqnarray}
&\displaystyle
\varepsilon_{m}\left(k_{\bot},k_{z}, \omega\right)=1-\frac{\omega^{2}_{pi}}{\omega^{2}}
+ \frac{1}{k^{2}\lambda^{2}_{De}}
\nonumber
\\  
&\displaystyle
\times
\left[ 1+ i\sqrt{\frac{\pi}{2}}\frac{\omega
-m\hat{\Omega}_{e}\left(R_{e0}\right)}{k_{z}
v_{Te}}W\left(z_{e}\right)A_{e0}\right]=0.
\label{38}
\end{eqnarray}
In Eq. (\ref{38}), $W\left(z_{e}\right)=e^{- z_{e}^{2}}\left(1 +\left(2i / \sqrt {\pi 
}\right)\int\limits_{0}^{z_{e}} e^{t^{2}}dt \right)$ is the Faddeeva function\cite{Faddeyeva} with 
argument $z_{e}=\left(\omega-m\Omega\left(R_{e0}\right)\right)/\sqrt{2}k_{z}v_{Te}$, $A_{e0}=I_{0}
\left(k^{2}_{\bot}\rho^{2}_{Te}\right)e^{-k^{2}_{\bot}\rho^{2}_{Te}}$,
and $I_{0}$ is the modified Bessel function of the first kind and order $0$. 
The frequency $\hat{\Omega}_{e}\left(R_{e0}\right)$, is determined as
\begin{eqnarray}
&\displaystyle
\hat{\Omega}_{e}\left(R_{e0}\right)=\Omega_{e}\left(R_{e0}\right)+\omega_{de}^{*}.
\label{39}
\end{eqnarray}
This frequency reveals the coupled effect of the plasma inhomogeneity, determined by the local 
electron diamagnetic drift frequency $\omega_{de}^{*}\left(R_{0e}\right)$,
\begin{eqnarray}
&\displaystyle
\omega_{de}^{*}\left(R_{0e}\right)=\omega_{ce}\rho^{2}_{Te}\frac{\partial \ln n_{0e}\left(R_{0e}
\right)}{R_{e0}\partial R_{e0}}\left(1-\frac{1}{2}\eta_{e}\right), 
\label{40}
\end{eqnarray}
where $\eta_{e}=\partial \ln T_{e}/\partial \ln n_{e0}$, of the radially inhomogeneous plasma 
with a cylindrical geometry, and the effect of the spatial inhomogeneity of the helicon wave 
field, determined by the frequency $\Omega_{e}\left(R_{e0}\right)$. It follows from Eqs. (\ref{20}) 
and (\ref{40}), that 
\begin{eqnarray}
&\displaystyle
\left|\frac{\omega_{de}^{*}}{\Omega_{e}\left(R_{e0}\right)}\right|\sim \frac{v^{2}_{Te}}
{\tilde{v}^{2}_{e}}\frac{\omega_{0}}{\left|\omega_{ce}\right|},
\label{41}
\end{eqnarray}
where
\begin{eqnarray}
&\displaystyle
\tilde{v}^{2}_{e}=\frac{c^{2}E_{1r}E_{1\varphi}}{B^{2}_{0}}.
\label{42}
\end{eqnarray}
Equation (\ref{41}) reveals that the effect of the plasma inhomogeneity is dominant in $
\hat{\Omega}_{e}\left(R_{e0}\right)$ when
\begin{eqnarray}
&\displaystyle
v_{Te}>\tilde{v}_{e}\left(\frac{\omega_{0}}{\left|\omega_{ce}\right|}\right)^{1/2}.
\label{43}
\end{eqnarray}
The opposite case of $\left|\Omega_{e}\left(R_{e0}\right)\right|>\left|\omega_{de}^{*}
\right|$ occurs when 
\begin{eqnarray}
&\displaystyle
\tilde{v}_{e}>v_{Te}\left(\frac{\left|\omega_{ce}\right|}{\omega_{0}}\right)^{1/2},
\label{44}
\end{eqnarray}
i. e. in the case of strong RF input power and a week magnetic field.

The solution to Eq. (\ref{38}) for the adiabatic electrons $\left(|z_{e}|\ll 1\right)$ is 
$\omega\left(\mathbf{k}\right)=\omega_{s}+\delta \omega\left(\mathbf{k}\right)$, where $
\omega_{s}\left(\mathbf{k}\right)$ is the frequency of the ion acoustic wave, $\omega^{2}_{s}
\left(\mathbf{k}\right)=k^{2}v^{2}_{s}\left(1+k^{2}\lambda^{2}_{De}\right)^{-1}$, 
$v_{s}=\left(T_{e}/m_{i}\right)^{1/2}$ is the ion acoustic velocity, and $\delta \omega\left(\mathbf{k}\right)$ 
with an accuracy to terms on the order of $\left(\delta \omega\left(\mathbf{k}\right)/\omega_{s}\right)^{2}
\ll 1$ is 
\begin{eqnarray}
&\displaystyle \delta\omega\left(\mathbf{k}\right)=-\frac{i\sqrt{\pi}}{2}\omega_{s}
\frac{z_{e0}}{\left(1+k^{2}\lambda^{2}_{De}\right)}W\left(z_{e0}\right)A_{e0}
\label{45}
\end{eqnarray}
with $z_{e0}=\left(\omega_{s}\left(\mathbf{k}\right)-m\hat{\Omega}
\left(R_{e0}\right)\right)/\sqrt{2}k_{z}v_{Te}$, where $|z_{e0}|< 1$ 
when $k_{z}/k>\sqrt{m_{e}/m_{i}}$. 
The ion acoustic instability develops when 
\begin{eqnarray}
&\displaystyle
m\hat{\Omega}\left(R_{e0}\right)>kv_{s},
\label{46}
\end{eqnarray}
with the growth rate $\gamma_{s}\left(\mathbf{k}\right)=\text{Im}\,\delta\omega\left(\mathbf{k}
\right)$ equal to 
\begin{eqnarray}
&\displaystyle \gamma_{s}\left(\mathbf{k}\right)=- \text{Im}\,\delta\omega\left(\mathbf{k}
\right) \approx \frac{\sqrt{\pi}}{2}\frac{\omega_{s}\left(\mathbf{k}\right)z_{e0}}{\left(1+k^{2}
\lambda^{2}_{De}\right)}e^{-z_{e0}^{2}}A_{e0}.
\label{47}
\end{eqnarray}
Note, that because $R_{e0}=m/k_{\bot}$, condition (\ref{46}) may be presented in the form  
$R_{e0}\hat{\Omega}\left(R_{e0}\right)>v_{s}$ i. e. the "azimuthal electron 
current velocity" should be larger than the ion acoustic velocity. The maximum growth rate (\ref{47})
attains for $z_{e0}=-1/\sqrt{2}$ and for $k_{\bot}\rho_{e}\gg 1$ it is equal to
\begin{eqnarray}
&\displaystyle \gamma_{max}\left(\mathbf{k}\right)\approx 0.054\frac{\left(\omega_{ce}\omega_{ci}
\right)^{1/2}}{\left(1+k^{2}\lambda^{2}_{De}\right)^{3/2}}.
\label{48}
\end{eqnarray}

It follows from Eqs. (\ref{38}), (\ref{39}), (\ref{40}) that by the transformations 
\begin{eqnarray}
&\displaystyle
\frac{m}{R_{e0}}\rightarrow k_{y}, \quad m\omega^{*}_{de}\rightarrow k_{y}v_{de}, 
\quad m\hat{\Omega}\rightarrow k_{y}V_{0\bot},
\label{49}
\end{eqnarray}
Eq. (\ref{38}) and its solutions (\ref{45}), (\ref{47}) becomes equal to the local dispersion 
equation for the slab model of the inhomogeneous plasma with electron current flowing 
perpendicularly to a magnetic field and to its solution for the ion acoustic current driven 
instability, respectively\cite{Lashmore-Davies}. By using this similarity of the considered 
microscale ion acoustic instability in the cylindrical and in the slab plasma geometries, we can 
employ the estimates for the energy density $W_{E}=\left(4\pi\right)^{-1}\int d\mathbf{k} k^{2}
\Phi^{2}\left(\mathbf{k}\right)$ of the electric field
\begin{eqnarray}
&\displaystyle
\frac{W_{E}}{n_{0e}T_{e}}\sim 5\cdot 10^{-4}\frac{\omega_{ce}}{\omega_{pe}}\frac{T_{e}}{T_{i}},
\quad \left(k\lambda_{De}\sim 1, \gamma=\gamma_{max}\right)
\label{50}
\end{eqnarray}
at the saturation state of the instability, resulted from the induced scattering of the ion acoustic 
wave by the unmagnetized ions\cite{Bychenkov}. The interaction of the magnetised electrons with ion 
acoustic turbulence under condition of the Cherenkov resonance results in the growth of the electron 
temperature $T_{e\parallel}$ along the magnetic field determined by the equation  
\begin{eqnarray}
&\displaystyle
n_{e}\frac{dT_{e\parallel}}{dt}\sim \frac{R_{e0}\hat{\Omega}\left(R_{e0}\right)}{v_{s}}
\left(\omega_{ci}\omega_{ce}\right)^{1/2}W_{E}
\nonumber
\\  
&\displaystyle \sim \nu_{eff}\frac{k^{2}_{0}c^{2}}{\omega^{2}_{pe}}W_{0}\left(R_{e0}\right),
\label{51}
\end{eqnarray}
where $W_{E}$ is determined by Eq. (\ref{50}), $W_{0}\left(R_{e0}\right)$ is the energy density of the helicon 
wave at radius $R_{e0}$ and $\nu_{eff}$ is the effective collision frequency of the electrons 
with electric field of the ion acoustic turbulence.

\section{Conclusions}\label{sec5}
In this paper, we present the theory of the microinstabilities of the cylindrical plasma 
excited by the cylindrically symmetric helicon wave with 
accounting for the cylindrical geometry and the radial inhomogeneities of the helicon wave 
and of a plasma. This theory reveals new macroscale effect of the azimuthal steady rotation of 
electrons with a radially inhomogeneous angular velocity, caused by the radial inhomogeneity of the 
helicon wave. It is found, that this effect is responsible for 
the development of the ion acoustic instability driven by the azimuthal electron current and plasma 
inhomogeneity at radius $R_{e0}$ where condition (\ref{46}) holds. In the vicinity of $R_{e0}$ 
this effect dominates over the effect of the oscillating  motion of the electrons relative to ions, 
which is basic in the theory of the parametric instabilities studied in Refs. \cite{Akhiezer, Aliev} 
with approximation of the spatially uniform helicon pumping wave.

\begin{acknowledgments}
This work was supported by National R\&D Program through the National Research Foundation of 
Korea (NRF) funded by the Ministry of Education, Science and Technology (Grant No. 
NRF-2017R1A2B2011106) and BK21 FOUR, the Creative Human Resource Education and Research 
Programs for ICT Convergence in the 4th Industrial Revolution, and by the National 
Research Foundation of Ukraine (Grant No.2020.02/0234).
\end{acknowledgments}

\bigskip
{\bf DATA AVAILABILITY}

\bigskip
The data that support the findings of this study are available from the corresponding author upon 
reasonable request.

\end{document}